\documentstyle[preprint]{ptptex}

\newcommand{\bra}[1]{\langle {#1} |}     
\newcommand{\ket}[1]{| {#1} \rangle}     
\newcommand{\bbra}[1]{\langle\!\langle {#1} |}     
\newcommand{\kket}[1]{| {#1} \rangle\!\rangle}     
\newcommand{\rket}[1]{| {#1} )}     
\newcommand{\maru}[1]{\breve{#1}} 

\title{
The $su(1,1)$-Algebraic Boson Model in \\
the Deformed Boson Scheme
}
\subtitle{
The Second Holstein-Primakoff Representation\\
as q-Deformed Boson Operator
}    

\author{
Atsushi {\sc Kuriyama},$^{1}$ 
Constan\c{c}a {\sc Provid\^encia}$^{2}$, \\
Jo\~ao da {\sc Provid\^encia}$^{2}$, Yasuhiko {\sc Tsue}$^{3}$ 
and Masatoshi {\sc Yamamura}$^{1}$
}

\inst{
$^1$Faculty of Engineering, Kansai University, Suita 564-8680, Japan\\
$^{2}$Departamento de Fisica, Universidade de Coimbra, P-3000 Coimbra, 
Portugal\\
$^{3}$Physics Division, Faculty of Science, Kochi University, Kochi 780-8520, 
Japan
}

\recdate{
}

\abst{
Following the deformed boson scheme leading to the $su(1,1)$-algebra, 
a certain simple boson system is deformed in the framework of the 
second Holstein-Primakoff representation. With the aid of the MYT boson 
mapping, the second representation arrives at the first Holstein-Primakoff 
representation. The $su(1,1)$-algebraic model obtained in this procedure 
is compared with that in the Schwinger representation investigated by 
three of the present authors (A.K., Y.T. and M.Y.). 
}

\begin{document}

\maketitle


The $q$-deformation provides us an attractive concept for understanding 
many-boson systems. However, if the quantity $[x]_q$ characterizing 
the deformation is restricted to the form $(q^x-q^{-x})/(q-q^{-1})$, 
variation of the deformations is not so rich as expected. 
Concerning the choice of $[x]_q$, Penson and Solomon stressed that 
the above form is nothing but one of the choices of the functions 
$[x]_q$.\cite{1} Recently, generalizing their viewpoint, 
the present authors proposed a possible method for the deformed boson scheme 
in the time-dependent variational method.\cite{2,3} Following this 
scheme, in Ref. \citen{3}, we formulated the $su(2)_q$- and the 
$su(1,1)_q$-algebras in the Holstein-Primakoff representation. 
Furthermore, in Ref. \citen{3}, we gave two simple applications mainly 
related to the $su(2)$-algebra. In this note, we present a simple 
application related to the $su(1,1)$-algebra, which enables us to describe 
the damped and the amplified oscillation in relation to thermal effects. 

Three of the present authors (A.K., Y.T. and M.Y.) investigated thermal 
effects appearing in various $su(2)$-algebraic models.\cite{4,5} 
The basis of this investigation is the use of the $su(1,1)$-algebraic 
boson model based on the Schwinger representation and the time-evolution 
is treated by the time-dependent variational method. 
The prototypes of this investigation are found in several papers.\cite{6,7} 
The time-evolutions of many-body systems, for example, the squeezed state, 
including the states discussed in Refs. \citen{4}, \citen{5} and \citen{7} 
are reviewed by the present authors.\cite{8} The aim of this note is to 
reinvestigate the same problem as that discussed in Refs. \citen{4} and 
\citen{5} in the frame of the $su(1,1)$-algebraic model in the 
Holstein-Primakoff representation, including some remarks on the 
results obtained in Ref. \citen{5} and the present paper.

For the above-mentioned aim, we adopt a many-boson system consisting 
of two kinds of boson operators $(\maru{c}, \maru{c}^*)$ and 
$(\maru{d}, \maru{d}^*)$. The Hamiltonian which concerns us is given 
in the form 
\begin{subequations}\label{1}
\begin{eqnarray}
& &\maru{H}=\maru{H}_c+\maru{H}_d+\maru{H}_{cd} \ , 
\label{1x}\\
& &\maru{H}_c=\epsilon \maru{c}^*\maru{c} \ , 
\label{1a}\\
& &\maru{H}_d=\omega \maru{d}^*\maru{d} \ , 
\label{1b}\\
& &\maru{H}_{cd}=-i\chi/2\cdot(\maru{c}^*\maru{d}-\maru{d}^*\maru{c}) \ . 
\label{1c}
\end{eqnarray}
\end{subequations}
Here, $\epsilon$, $\omega$ and $\chi$ denote real parameters. In associating 
with the form (\ref{1a}), we note the following relation : 
\begin{subequations}\label{2}
\begin{eqnarray}
& &\maru{c}^*\maru{c}
=\sqrt{\frac{2(1+\zeta)}{1+[\maru{c} , \maru{c}^* ] +2\zeta}}\ 
\maru{c}^* \maru{c} \ 
\sqrt{\frac{2(1+\zeta)}{1+[\maru{c} , \maru{c}^* ] +2\zeta}} \ .
\label{2a}
\end{eqnarray}
Here, $\zeta$ denotes a real parameter and, later, it is regarded 
as infinitesimal. In Ref.\citen{3}, we used, in the present notation, 
\begin{equation}
\maru{c}^* \maru{c}=\maru{c}^* 
\frac{2}{1+[ \maru{c} , \maru{c}^* ]}\maru{c} \ .
\label{2b}
\end{equation}
\end{subequations}
Since $[\maru{c},\maru{c}^*]=1$, the relation (\ref{2a}) and (\ref{2b}) 
are identical equations.

Following the basic idea shown in the relation (30) of Ref. \citen{3}, 
let the boson $(\maru{c}, \maru{c}^*)$ deform in the form 
\begin{eqnarray}\label{3}
& &\maru{\gamma}=\sqrt{\maru{N}/n_0-1}\ \maru{c}
=(\sqrt{n_0})^{-1}\sqrt{\maru{N}-n_0} \ \maru{c} \ , \nonumber\\
& &\maru{\gamma}^*=\maru{c}^*\ \sqrt{\maru{N}/n_0-1}
=(\sqrt{n_0})^{-1}\maru{c}^*\ \sqrt{\maru{N}-n_0} \ , \nonumber\\
& &\maru{N}=\maru{c}^*\maru{c} \ . 
\end{eqnarray}
Here, $n_0$ denotes a positive integer. The commutation relation of 
$\maru{\gamma}$ and $\maru{\gamma}^*$ is obtained in the form 
\begin{equation}\label{4}
[ \maru{\gamma} , \maru{\gamma}^* ]=2\maru{N}/n_0-1 \ .
\end{equation}
The second Holstein-Primakoff representation of the $su(1,1)$-algebra, 
which was called by the present authors in Ref. \citen{3}, is given as 
\begin{equation}\label{5}
\maru{T}_-=\sqrt{n_0}\cdot\maru{\gamma}\ , \qquad
\maru{T}_+=\sqrt{n_0}\cdot\maru{\gamma}^*\ , \qquad
\maru{T}_0=(n_0/2)\cdot [\maru{\gamma} , \maru{\gamma}^* ] \ . 
\end{equation}
We do not deform the boson $(\maru{d}, \maru{d}^*)$. 
Under the above scheme, we perform the deformation for the Hamiltonian 
(\ref{1x}), which is formally written as 
\begin{equation}\label{6}
\maru{H}^{\rm (def)}=\maru{H}_c^{\rm (def)}+\maru{H}_d^{\rm (def)} 
+\maru{H}_{cd}^{\rm (def)} \ . 
\end{equation}
For the deformation of $\maru{H}_c$, the relation (\ref{2a}) is useful. 
In the present case, 
the relation (\ref{2a}) leads us to 
\begin{equation}\label{7}
\sqrt{\frac{2(1+\zeta)}{1+[\maru{\gamma} , \maru{\gamma}^* ] +2\zeta}}\ 
\maru{\gamma}^* \maru{\gamma} \ 
\sqrt{\frac{2(1+\zeta)}{1+[\maru{\gamma} , \maru{\gamma}^* ] +2\zeta}} 
=[\maru{c}^*\maru{c}-(n_0+1)]\cdot \frac{(1+\zeta)\maru{N}}{\maru{N}+\zeta} \ .
\end{equation}
As is clear from the relation (\ref{3}), $({\hat N}/n_0-1)$ should 
be positive definite and, then, at the limit $\zeta\rightarrow 0$, 
$(1+\zeta)\maru{N}/(\maru{N}+\zeta)\rightarrow 1$. 
Thus, we have 
\begin{equation}\label{8}
[\maru{c}^*\maru{c}-(n_0+1)]\cdot \frac{(1+\zeta)\maru{N}}{\maru{N}+\zeta} 
\longrightarrow \maru{c}^*\maru{c}-(n_0+1) \ .
\end{equation}
Then, $\maru{H}_c^{\rm (def)}$ is written in the form 
\begin{subequations}\label{6all}
\begin{equation}
\maru{H}_c^{\rm (def)}=\epsilon \maru{c}^*\maru{c} -\epsilon(n_0+1) \ .
\label{6a}
\end{equation}
As was already mentioned, we do not deform the boson $(\maru{d}, \maru{d}^*)$. 
Therefore, we have 
\begin{eqnarray}
& &\maru{H}_d^{\rm (def)}=\omega\maru{d}^*\maru{d} \ , 
\label{6b}\\
& &\maru{H}_{cd}^{\rm (def)}
=-i(\chi/2\sqrt{n_0})\cdot[\maru{c}^*\sqrt{\maru{N}-n_0}\ \maru{d} 
-\maru{d}^*\sqrt{\maru{N}-n_0}\ \maru{c} ] \ .
\label{6c}
\end{eqnarray}
\end{subequations}
The above is the present deformed Hamiltonian. 
The term, $-\epsilon(n_0+1)$, in $\maru{H}_c^{({\rm def})}$ plays 
a role of shifting the value of the energy and does not give any 
influence on the dynamics.

It may be self-evident that the Hamiltonian (\ref{6}) is treated in the 
orthogonal set\break
$\{ \kket{n_0+1+n} \ ; \ n=0,1,2,\cdots \}$ : 
\begin{equation}\label{9}
\kket{n_0+1+n}=\left(\sqrt{(n_0+1+n)!}\right)^{-1}\cdot
(\maru{c}^*)^{n_0+1+n} \kket{0} \ .
\end{equation}
Clearly, concerning the boson number $\maru{N}\ (=\maru{c}^*\maru{c})$, 
the minimum value is $(n_0+1)$. Therefore, it may be convenient to 
describe the boson number in terms of the minimum value plus its 
fluctuation. For this aim, the idea of the MYT boson mapping\cite{9} 
is available. The mapped space consists of the orthogonal set 
$\{ \ket{n}\ ; \ n=0,1,2,\cdots \}$ : 
\begin{equation}\label{10}
\ket{n}=(\sqrt{n!})^{-1}\cdot ({\hat \alpha}^*)^n \ket{0} \ . 
\qquad ({\hat \alpha}\ket{0}=0) 
\end{equation}
Of course, $(\hat{\alpha}, \hat{\alpha}^*)$ denotes the boson operator. 
Then, we define the mapping operator $U$ in the form 
\begin{subequations}\label{11all}
\begin{eqnarray}
& &U=\sum_{n=0}^{\infty} \ket{n}\bbra{n_0+1+n} \ , 
\label{11}\\
& &U U^{\dagger}=\sum_{n=0}^{\infty} \ket{n}\bra{n}=1 \ , 
\nonumber\\
& &U^{\dagger}U=\sum_{n=0}^{\infty}\kket{n_0+1+n}\bbra{n_0+1+n}
=1-\sum_{k=0}^{n_0} \kket{k}\bbra{k} \ . 
\label{11b}
\end{eqnarray}
\end{subequations}
The operator $U$ presents us the relation 
\begin{equation}\label{12}
U\kket{n_0+1+n}=\ket{n} \ .
\end{equation}
The operator $\maru{c}^*\maru{c}$ is mapped on the form 
\begin{equation}\label{13}
U\maru{c}^*\maru{c} U^{\dagger} =n_0+1+\hat{N} \ , \qquad
\hat{N}=\hat{\alpha}^*\hat{\alpha} \ .
\end{equation}
Further, for the set $\{ T_- , T_+, T_0 \}$, we have 
\begin{eqnarray}\label{14}
& &U\maru{T}_- U^{\dagger} =\hat{T}_- \ , \qquad
{\hat T}_-=\sqrt{n_0+2+{\hat  N}}\ \hat{\alpha} \ , \nonumber\\
& &U\maru{T}_+ U^{\dagger} =\hat{T}_+ \ , \qquad
{\hat T}_+=\hat{\alpha}^*\sqrt{n_0+2+{\hat  N}} \ , \nonumber\\
& &U\maru{T}_0 U^{\dagger} =\hat{T}_0 \ , \qquad
{\hat T}_0=(n_0+2)/2+{\hat  N}\ . 
\end{eqnarray}
It may be clear that the form (\ref{13}) comes up to our expectation. 
The operator $\hat{N}$ describes the fluctuation. The set 
$\{ {\hat T}_- , \hat{T}_+ , \hat{T}_0 \}$ denotes the first 
Holstein-Primakoff representation of the $su(1,1)$-algebra, 
in which the magnitude of the $su(1,1)$-spin is $(n_0+2)/2$. 
This form was also discussed in Ref. \citen{3}. 
Thus, the Hamiltonian mapped from the form (\ref{6}) can be 
expressed as follows : 
\begin{subequations}\label{15all}
\begin{eqnarray}
& &{\hat H}^{\rm (def)}={\hat H}_c^{\rm (def)}+{\hat H}_d^{\rm (def)} 
+\hat{H}_{cd}^{\rm (def)} \ , 
\label{15}\\
& &{\hat H}_c^{\rm (def)}=\epsilon (n_0+1)
+\epsilon {\hat \alpha}^*{\hat \alpha} \ , 
\label{15a}\\
& &{\hat H}_d^{\rm (def)}=\omega \maru{d}^*\maru{d} \ , 
\label{15b}\\
& &{\hat H}_{cd}^{\rm (def)}=-i(\chi/2\sqrt{n_0})\cdot
[{\hat \alpha}^*\sqrt{n_0+2+\hat{N}}\ \maru{d}-\maru{d}^*
\sqrt{n_0+2+\hat{N}}\ {\hat \alpha}] \ .
\label{15c}
\end{eqnarray}
\end{subequations}

For the Hamiltonian (\ref{15}), the following picture can be drawn : 
Let us imagine that 
the external environment, such as heat bath, is described by the 
boson $(\maru{d} , \maru{d}^*)$. The harmonic oscillation described by 
the boson $(\maru{c} , \maru{c}^*)$ may be damped or amplified in terms of 
the interaction to the external environment which is assumed to be 
extremely big system. Therefore, for the boson $(\maru{d} , \maru{d}^*)$, 
we can neglect the fluctuation around the equilibrium value. 
This means that $(\maru{d} , \maru{d}^*)$ can be replaced by the 
time-independent $c$-number $(\delta , \delta^*)$. It may be 
performed by calculating the expectation value of $\hat{H}^{\rm (def)}$ 
for the boson coherent state for $(\maru{d} , \maru{d}^*)$. 
The $c$-number $(\delta , \delta^*)$ can be expressed as 
\begin{equation}\label{16}
\delta=|\delta|\exp(-i\phi) \ , \qquad 
\delta^*=|\delta|\exp(+i\phi) \ . 
\end{equation}
Then, the boson $(\hat{\alpha} , \hat{\alpha}^*)$ is redefined in the form 
\begin{equation}\label{17}
{\hat \alpha} \longrightarrow {\hat \alpha}\exp(+i\phi) \ , 
\qquad
{\hat \alpha}^* \longrightarrow {\hat \alpha}^*\exp(-i\phi) \ . 
\end{equation}
With the use of the relation (\ref{16}) and (\ref{17}), the 
Hamiltonian (\ref{15}) can be modified in the following form : 
\begin{subequations}\label{18all}
\begin{eqnarray}
& &{\hat H}=E+{\hat H}_0+\hat{H}_{i} \ , 
\label{18}\\
& &E=\epsilon (n_0+1)+\omega|\delta|^2 \ , 
\label{18a}\\
& &{\hat H}_0=\epsilon \hat{\alpha}^*\hat{\alpha}
=-\epsilon (n_0+2)/2+\epsilon {\hat T}_0 \ , 
\label{18b}\\
& &{\hat H}_{i}=-i(\eta/2)\!\cdot\!
[{\hat \alpha}^*\sqrt{n_0+2+\hat{N}}-\sqrt{n_0+2+\hat{N}}\ {\hat \alpha}]
=-i(\eta/2)\!\cdot\! (\hat{T}_+-\hat{T}_-) \ , \qquad
\label{18c}\\
& &\eta=\chi|\delta|/\sqrt{n_0} \ . 
\label{19}
\end{eqnarray}
\end{subequations}
The Hamiltonian (\ref{18}) is essentially of the same form as that derived 
in nuclear $su(2)$-models in the Schwinger boson representation for the 
$su(1,1)$-algebra.\cite{4} 
The Schwinger representation for the $su(1,1)$-algebra is given as 
\begin{equation}\label{14a}
{\hat T}_-^{(s)}={\hat a}{\hat b} \ , \qquad
{\hat T}_+^{(s)}={\hat b}^*{\hat a}^* \ , \qquad
{\hat T}_0^{(s)}=({\hat b}{\hat b}^*+{\hat a}^*{\hat a})/2 \ . 
\end{equation}
Here, $({\hat a} , {\hat a}^*)$ and $({\hat b} , {\hat b}^*)$ 
denote boson operators. If $({\hat T}_- , {\hat T}_+ , {\hat T}_0)$ 
in the Hamiltonian (\ref{18}) is replaced with the form (\ref{14a}), 
it is reduced to the Hamiltonian derived in Ref. \citen{4}. Further, the 
time-evolution of the system described in terms of the Hamiltonian, which 
is essentially the same as that shown in the form (\ref{18}), was 
investigated in the framework of the time-dependent variational 
method.\cite{5} The prototypes of these investigations can be found in 
several papers.\cite{6,7} The time-evolution of many-body systems 
including the problems discussed in Refs. \citen{6} and \citen{7} 
are reviewed by the present authors.\cite{8}

As a trial state for the variation, in Ref. \citen{5}, the following 
wave packet was adopted : 
\begin{subequations}\label{20all}
\begin{eqnarray}
& &\rket{c}=U^{-1}\exp(-|W|^2)\cdot\exp[(V/U){\hat T}_+^{(s)}]\rket{m} \ , 
\label{20}\\
& &\rket{m}=\exp(\sqrt{2}W/U\cdot {\hat b}^*)\rket{0} \ . 
\qquad ({\hat a}\rket{0}={\hat b}\rket{0}=0 \ , \quad((c\rket{c}=0) 
\label{20a}
\end{eqnarray}
\end{subequations}
Here, $U$ and $V$ denote real and complex parameters, respectively, 
satisfying 
\begin{equation}\label{21}
U^2-|V|^2=1 \ .
\end{equation}
The quantity $W$ is also complex and $(2|W|^2+1)/2$ plays a role of the 
magnitude of the $su(1,1)$-spin, such as $(n_0+2)/2$. 
It should be noted that the state $\rket{m}$ obeys the condition 
\begin{equation}\label{22}
{\hat T}_-^{(s)}\rket{m}=0 \ .
\end{equation}
The wave packet (\ref{20}) is translated into the following form, 
$\ket{c}$, in the case of the Holstein-Primakoff representation for 
the $su(1,1)$-algebra : 
\begin{equation}\label{23}
\ket{c}=u^{-(n_0+2)}\cdot\exp[(v/u)\hat{\alpha}^*\sqrt{n_0+2+\hat{N}}]
\ket{0} \ . \qquad (\bra{c}c\rangle=1)
\end{equation}
Here, $u$ and $v$ denote real and complex parameters, respectively, 
obeying 
\begin{equation}\label{24}
u^2-|v|^2=1 \ . 
\end{equation}
In the case of the Schwinger representation, there exist infinite 
possibilities for selecting the state $\rket{m}$ and, in Ref. \citen{5}, 
the form (\ref{20a}) was adopted. 
However, in the present case, the state obeying the condition (\ref{22}) is 
uniquely given as $\ket{m}=\ket{0}$.

The time-dependence of $v$ can be determined through the time-dependent 
variational method for the system under investigation. 
The state $\ket{c}$ is rewritten in the following form : 
\begin{eqnarray}
& &\ket{c}=\sum_{n=0}^{\infty}
\sigma_n \ket{n} \ , \qquad (\sum_{n=0}^{\infty}|\sigma_n|^2=1) 
\label{25}\\
& &\sigma_n=(v^n/u^{n_0+2+n})\sqrt{(n_0+1+n)!/n!(n_0+1)!} \ .
\label{26}
\end{eqnarray}
The state $\ket{n}$ is given in the relation (\ref{10}). 
With the use of the canonicity condition,\cite{10} 
$(v, v^*)$ can be parametrized in terms of the canonical variable 
$(\alpha , \alpha^*)$ in the boson type which obeys the Poisson 
bracket relation 
\begin{equation}\label{27}
[ \alpha , \alpha^* ]_P=-i \ .
\end{equation}
The parameter $(v , v^*)$ is given by 
\begin{equation}\label{28}
v=\left(\sqrt{n_0+2}\right)^{-1}\cdot \alpha \ , \qquad
v^*=\left(\sqrt{n_0+2}\right)^{-1}\cdot \alpha^* \ . 
\end{equation}
The expectation values of the operators composing the Hamiltonian (\ref{18}) 
are obtained as 
\begin{eqnarray}
& &\bra{c}{\hat T}_-\ket{c}=(n_0+2)uv=\sqrt{n_0+2+N}\ \alpha \ , 
\nonumber\\
& &\bra{c}{\hat T}_+\ket{c}=(n_0+2)uv^*=\alpha^*\sqrt{n_0+2+N} \ , 
\nonumber\\
& &\bra{c}{\hat T}_0\ket{c}=(n_0+2)/2+(n_0+2)|v|^2=
(n_0+2)/2+N , 
\label{29}\\
& &N=\alpha^*\alpha \ . 
\label{30}
\end{eqnarray}
We can see that, in the Poisson bracket relation (\ref{27}), the 
expectation value (\ref{29}) is classical counterpart of 
$({\hat T}_- , {\hat T}_+ , {\hat T}_0)$. Under the above preparation, 
the Hamilton equation for the expectation value $\bra{c}{\hat H}\ket{c}$ 
as the classical Hamiltonian can be solved and the solutions are 
classified into the following three groups : 
\begin{subequations}\label{31all}
\begin{eqnarray}
& &\hbox{\rm 1)}\ \epsilon^2 > \eta^2 \ : \ 
\hbox{\rm the\ solution\ is\ expressed\ in\ terms\ of\ trigometric\ 
function\ for\ } \nonumber\\
& &\qquad\qquad\qquad
\sqrt{\epsilon^2-\eta^2}\ t \ , 
\label{31a}\\
& &\hbox{\rm 2)}\ \epsilon^2 = \eta^2 \ : \ 
\hbox{\rm the\ solution\ is\ expressed\ in\ terms\ of\ quadratic\ 
function\ for\ } \nonumber\\
& &\qquad\qquad\qquad
\epsilon t \ , 
\label{31b}\\
& &\hbox{\rm 3)}\ \epsilon^2 < \eta^2 \ : \ 
\hbox{\rm the\ solution\ is\ expressed\ in\ terms\ of\ hyperbolic\ 
function\ for\ } \nonumber\\
& &\qquad\qquad\qquad
\sqrt{\eta^2-\epsilon^2}\ t \ .  
\label{31c}
\end{eqnarray}
\end{subequations}
The details are found in Refs. \citen{5} and \citen{8}.

In Ref. \citen{5}, thermal effects observed in the $su(1,1)$-algebraic 
model were investigated. Of course, the basic idea comes from the 
prototype of the investigation.\cite{7} The viewpoint was as follows : 
Presupposing that the wave packet (\ref{20}) expresses the statistically 
mixed state, the entropy can be introduced in the present system and, 
various quantities characterizing thermal effects can be calculated. 
In this paper, we borrow the above viewpoint. If we stand on this point, 
the quantity $|\sigma_n|^2$ appearing in the relations (\ref{25}) and 
(\ref{26}) denotes the statistically mixed weight for the pure state 
labeled by $n$. Then, the entropy $\sigma$ in the standard form is 
given as 
\begin{equation}\label{32}
\sigma=\sum_{n=0}^{\infty} |\sigma_n|^2\cdot \log |\sigma_n|^2 \ . 
\end{equation}
Of course, $|\sigma_n|^2$ is given as 
\begin{equation}\label{33}
|\sigma_n|^2=[(|v|^2)^n/(u^2)^{n_0+2+n}]\cdot 
[(n_0+1+n)!/n!(n_0+1)!] \ . 
\end{equation}
In principle, we can calculate the entropy directly by the form 
(\ref{32}). However, it may be impossible to provide a systematic 
method for the calculation of the part of the logarithm. 
Then, we borrow the form presented by Tsallis\cite{11} as a 
tool for the calculation of the entropy (\ref{32}). 
With the relation to the standard form (\ref{32}), Tsallis' 
form is given as 
\begin{eqnarray}
& &\sigma(q)=\left( \sum_{n=0}^{\infty} (|\sigma_n|^2)^q-1\right)/(1-q) \ . 
\qquad (q\ : \ \hbox{\rm possitive\ parameter}) 
\label{34}\\
& &\sigma=\lim_{q\rightarrow 1} \sigma(q) \ . 
\label{35}
\end{eqnarray}
For the calculation of $\sigma(q)$ in terms of the power series expansion 
for the parameter $(n_0+1)$, the following formula for $(n_0+1)$ 
is useful : 
\begin{subequations}\label{36all}
\begin{equation}\label{36}
[(n_0+1+n)!/n!(n_0+1)!]^q=1+\sum_{k=1}^{\infty}
f_k(n)/n!\cdot (n_0+1)^k \ . \quad 
(n\ge 1)
\end{equation}
Here, $f_k(n)$ obeys the recursion formula 
\begin{equation}\label{36a}
f_{k+1}(n)=q\sum_{m=0}^{k}(-)^m[k!/(k-m)!]\cdot f_{k-m}(n)\sum_{r=1}^{n}
r^{-(m+1)} \ . \quad (f_0(n)=1) 
\end{equation}
For example, we have 
\begin{eqnarray}\label{36b}
& &f_1(n)=q\left(\sum_{r=1}^n r^{-1}\right) \ , 
\qquad
f_2(n)=q^2\left(\sum_{r=1}^n r^{-1}\right)^2-q\left(\sum_{r=1}^n r^{-2}\right) 
\ , \nonumber\\
& &f_3(n)=q^3\left(\sum_{r=1}^n r^{-1}\right)^3 
-3q^2\left(\sum_{r=1}^n r^{-1}\right)\left(\sum_{r=1}^n r^{-2}\right)
+2q\left(\sum_{r=1}^n r^{-3}\right) \ . 
\end{eqnarray}
\end{subequations}
Straightforward calculation based on the formula (\ref{36}) leads us to 
the form 
\begin{eqnarray}\label{37}
\sum_{n=0}^{\infty}(|\sigma_n|^2)^q
&=&[(u^2)^q-(|v|^2)^q]^{-1} \nonumber\\
& &\times \Bigl\{1+(n_0+1)q[(q-1)\log u^2-\log [(u^2)^q-(|v|^2)^q]] \nonumber\\
& &\ \ \ \ +(n_0+1)^2/2\cdot q^2[(q-1)\log u^2-\log [(u^2)^q-(|v|^2)^q]]^2 
\nonumber\\
& &\ \ \ \ +(n_0+1)^2/2\cdot q(q-1)\sum_{r=1}^{\infty}[(|v|^2/u^2)^q]^r
\cdot r^{-2} + \cdots \Bigl\} \ . 
\end{eqnarray}
It is enough for obtaining the limit (\ref{35}) to calculate 
$[(u^2)^q-(|v|^2)^q]^{-1}$ and 
$\log [(u^2)^q-(|v|^2)^q]$ up to the order of $(q-1)$ : 
\begin{eqnarray}\label{38}
& &[(u^2)^q-(|v|^2)^q]^{-1}=1-(q-1)(u^2\log u^2-
|v|^2 \log |v|^2 ) \ , 
\nonumber\\
& &\log [(u^2)^q-(|v|^2)^q]=(q-1)(u^2 \log u^2-|v|^2 \log |v|^2) \ . 
\end{eqnarray}
Substituting the relation (\ref{38}) into the form (\ref{37}) and 
calculating the limit (\ref{35}), we have the following form for the 
entropy : 
\begin{subequations}\label{39all}
\begin{eqnarray}
& &\sigma=\sigma_0+(n_0+1)\cdot \sigma_1+(n_0+1)^2/2\cdot \sigma_2+\cdots \ , 
\label{39}\\
& &\sigma_0=u^2\cdot \log u^2-|v|^2\cdot\log |v|^2 \ , 
\label{39a}\\
& &\sigma_1=|v|^2\cdot \log u^2-|v|^2\cdot\log |v|^2 \ , 
\label{39b}\\
& &\sigma_2=-(|v|^2\cdot \log u^2-|v|^2\cdot\log |v|^2)
-\sum_{r=1}^{\infty} (|v|^2/u^2)^r\cdot r^{-2} \ . 
\label{39c}
\end{eqnarray}
\end{subequations}
The above is expressed in the frame of the quadratic for $(n_0+1)$. 
On the other hand, the entropy calculated in the wave packet 
(\ref{20}) is given in the form\cite{8} 
\begin{subequations}\label{40all}
\begin{eqnarray}
& &\sigma^{(s)}
=\sigma_0^{(s)}+(2|W|^2)\cdot \sigma_1^{(s)}
+(2|W|^2)^2/2\cdot \sigma_2^{(s)}+\cdots \ , 
\label{40}\\
& &\sigma_0^{(s)}=U^2\cdot \log U^2-|V|^2\cdot\log |V|^2 \ , 
\label{40a}\\
& &\sigma_1^{(s)}=|V|^2\cdot \log U^2-|V|^2\cdot\log |V|^2 \ , 
\label{40b}\\
& &\sigma_2^{(s)}=-|V|^2/U^2 \ . 
\label{40c}
\end{eqnarray}
\end{subequations}
The two wave packets (\ref{20}) and (\ref{23}) are different from each 
other and we can see that the difference in the expression of the entropy 
appears in the third term. In Ref. \citen{5}, the third term was neglected 
under the expectation that this term may be small in a rather wide region 
of $|V|^2$.\cite{5,8} 
If this expectation is conserved in the present treatment, we can neglect 
the third term and the result is completely identical to that 
in Ref. \citen{5}. 
Thus, in the frame of the Holstein-Primakoff representation, we can reproduce 
the results obtained in the Schwinger representation.

However, the term in the expression of the entropy (\ref{39}) 
can be neglected or not ? The region, in which the third term is neglected, 
seems to be not so wide as that expected in Ref. \citen{5}. 
The impromptu idea for saving this situation is the use of the 
Pad\'e approximation. In this case, the expression (\ref{39}) is modified in 
the form 
\begin{equation}\label{41}
\sigma=\sigma_0\cdot [1-(n_0+1)(\sigma_2/2\sigma_1-\sigma_1/\sigma_0)]
[1-(n_0+1)(\sigma_2/2\sigma_1)]^{-1} \ .
\end{equation}
The form (\ref{41}) tells us the following : 
In the small region of $(n_0+1)$, the main part comes from $\sigma_0$. 
On the other hand, in the large region, the main term appears in the 
form $\sigma_0(1-2\sigma_1^2/\sigma_0\sigma_2)$ and the term 
$2\sigma_1^2/\sigma_2$ may be of the same order. 
From the above simple consideration, the form (\ref{41}) is expected to 
show rather stable behavior. Of course, for the calculation of the 
form (\ref{41}), the explicit expression of the function 
$f(x)=\sum_{r=1}^{\infty} x^r\cdot r^{-2}$ is necessary, but, it is 
impossible to express it in terms of any elementary functions. 
A possible approximate expression in the region $0\le x \le 1$, 
which is not so popular, is given, for example, in the form 
\begin{subequations}\label{42all}
\begin{eqnarray}\label{42}
f(x)&\sim& 1+\sum_{n=1}^{m-1} [n^2(n+1)]^{-1}x^n
+(m+1)[m^2(2m+1)]^{-1}x^m \nonumber\\
& &\ \ +(1-x)/x\cdot \log(1-x) \ . 
\end{eqnarray}
Here, $m$ denotes positive integer $(m=1,2,3,\cdots)$. 
If $m$ is chosen as large, the approximation becomes better. 
In the case $m=4$, we have 
\begin{eqnarray}\label{42a}
f(x)&\sim& 1+(1/2)\cdot x+(1/12)\cdot x^2+(1/36)\cdot x^3
+(5/144)\cdot x^4 \nonumber\\
& &\ \ +(1-x)/x\cdot\log(1-x) \ .
\end{eqnarray}
\end{subequations}
For example, we know the exact values $f(1/2)=\pi^2/12-(\log 2)^2/2\ 
(=0.5822)$ and $f(1)=\pi^2/6\ (=1.6449)$. The form (\ref{42a}) gives 
us the approximate values $f(1/2)\sim 0.5833$ and $f(1)\sim 1.6458$.

Finally, we give three remarks. First is related with the use of the 
state $\ket{c}$ shown in the relation (\ref{20}). For the present formalism, 
the use of the state (\ref{20}) is reasonable or not ? It should be discussed 
in relation to another form investigated by the present authors.\cite{12} 
As the second remark, we should note that in the present treatment, 
it is impossible to describe the cases $n_0=0$ and 1. It may be 
self-evident that the reason comes from the form (\ref{3}). 
Third is as follows : The state $\ket{c}$ can be regarded as expressing 
the statistically mixed state or not ? In the case of the Schwinger 
representation, the degree of freedom characterizing the statistically 
mixed state is included in terms of the boson $({\hat a}, {\hat a}^*)$ 
in addition to the degree of freedom for describing the physical 
object by the boson $({\hat b}, {\hat b}^*)$. 
The present formalism does not contain such degree of freedom explicitly. 
Including such questions, we will contact with the problem discussed 
in this paper in more details.

\section*{Acknowledgements}
\vspace{-0.2cm}

The main part of this work was performed when two of the authors 
(Y.T. and M.Y.) stayed at Coimbra in August and September of 2002. 
They express their sincere thanks to Professor J. da Provid\^encia, 
co-author of this paper.

\end{document}